\documentclass[a4paper,fleqn,usenatbib]{mnras}

\usepackage{newtxtext,newtxmath}

\usepackage[T1]{fontenc}
\usepackage{ae,aecompl}

\usepackage[usenames,dvipsnames,svgnames]{xcolor}
\usepackage{ulem}
\usepackage{longtable}

\usepackage{graphicx}
\usepackage{lscape}
\usepackage{xfrac}

\newcommand*{\bfrac}[2]{\genfrac{\lbrace}{\rbrace}{0pt}{}{#1}{#2}}

\def\ergscm{erg~s$^{-1}$~cm$^{-2}$}
\def\arcdeg{\hbox{$^\circ$}}
\def\arcmin{\hbox{$^\prime$}}

\def\lum{erg s$^{-1}$}

\def\aj{AJ}
\def\apj{ApJ}

\def\aap{A\&A}
\def\mnras{MNRAS}
\def\apjs{ApJS}

\def\nat{Nature}
\def\apjl{ApJLett}
\def\procspie{\ref@jnl{Proc.~SPIE}}   

\def\int{\textit{INTEGRAL}}
\def\nu{\textit{NuSTAR}}

\newcommand*{\mysim}{\mathord{\sim}}

\title[New INTEGRAL sources in the GP after 14 years]{New hard X-ray sources discovered in the ongoing INTEGRAL Galactic Plane survey after 14 years of observations}
\author[Krivonos et al.]{Roman
  A. Krivonos,$^{1}$\thanks{E-mail: krivonos@iki.rssi.ru} Sergey
  S. Tsygankov,$^{2,1}$ Ilya A. Mereminskiy,$^1$ \newauthor Alexander
  A. Lutovinov,$^{1,3}$ Sergey Yu. Sazonov$^{1,3}$ and Rashid A. Sunyaev$^{4,1}$
\\
$^{1}$Space Research Institute of the Russian Academy of Sciences,
Profsoyuznaya Str. 84/32, 117997 Moscow, Russia\\
$^{2}$Tuorla Observatory, Department of Physics and Astronomy,
University of Turku, V\"ais\"al\"antie 20, FI-21500 Piikki\"o, Finland\\
$^{3}$Moscow Institute of Physics and Technology, Institutsky per. 9, 141700 Dolgoprudny, Russia\\
$^{4}$MPI f\"ur Astrophysik, Karl-Schwarzschild str. 1, Garching
D-85741, Germany }

\begin{document}
\label{firstpage}
\pagerange{\pageref{firstpage}--\pageref{lastpage}}
\maketitle

\begin{abstract}
  The International Gamma-Ray Astrophysics Laboratory (INTEGRAL)
  continues to successfully work in orbit after its launch in
  2002. The mission provides the deepest ever survey of hard X-ray
  sources throughout the Galaxy at energies above 20 keV. We report on
  a catalogue of new hard X-ray source candidates based on the latest
  sky maps comprising 14 years of data acquired with the IBIS
  telescope onboard \int\ in the Galactic Plane
  ($|b|<17.5$\arcdeg). The current catalogue includes in total {72}
  hard X-ray sources detected at $S/N>4.7\sigma$ and not known to
  previous \int\ surveys. Among them, {31} objects have also been
  detected in the on-going all-sky survey by the BAT telescope of the
  \textit{Swift} observatory. For {26} sources on the list, we suggest
  possible identifications: {21} active galactic nuclei, {two}
  cataclysmic variables, {two} isolated pulsars or pulsar wind
  nebulae, and one supernova remnant; {46} sources from the catalogue
  remain unclassified.
\end{abstract}


\section{Introduction}

X-ray surveys play a key role in our understanding of energetic
phenomena in the Universe. Detailed investigations of the physics and
evolution of X-ray selected sources are usually based on
systematic studies of their properties. Observations in recent
decades have revealed a variety of X-ray point sources beyond the
solar system in the Milky Way and Magellanic Clouds. Although the
bright X-ray sources in the Milky Way can be effectively studied, many
of them are not observable due to the heavy obscuration by the
Galactic disk. Studies of nearby galaxies with modern sensitive soft
X-ray telescopes are relatively free from the obscuration problem and can
provide us uniform samples of X-ray binaries in different environments
\citep[see][for a review]{2006csxs.book..475F,2006ARA&A..44..323F}. As
a result, we may know better the properties of X-ray source
populations and structure of the nearby galaxies, than of our own
Milky Way.

X-ray observations of our Galaxy at energies above 10 keV are free
from the obscuration bias. However, due to the large extent of the Milky
Way across the sky, a systematic survey of the Galactic X-ray source
population and discovery of new X-ray emitters require wide-angle
instruments. This makes the IBIS coded-mask telescope
\citep{2003A&A...411L.131U} onboard the \int\ observatory
\citep{2003A&A...411L...1W} unique and most suitable for surveying the
Galaxy in the hard X-ray domain.

\begin{figure}
\includegraphics[width=1.05\columnwidth]{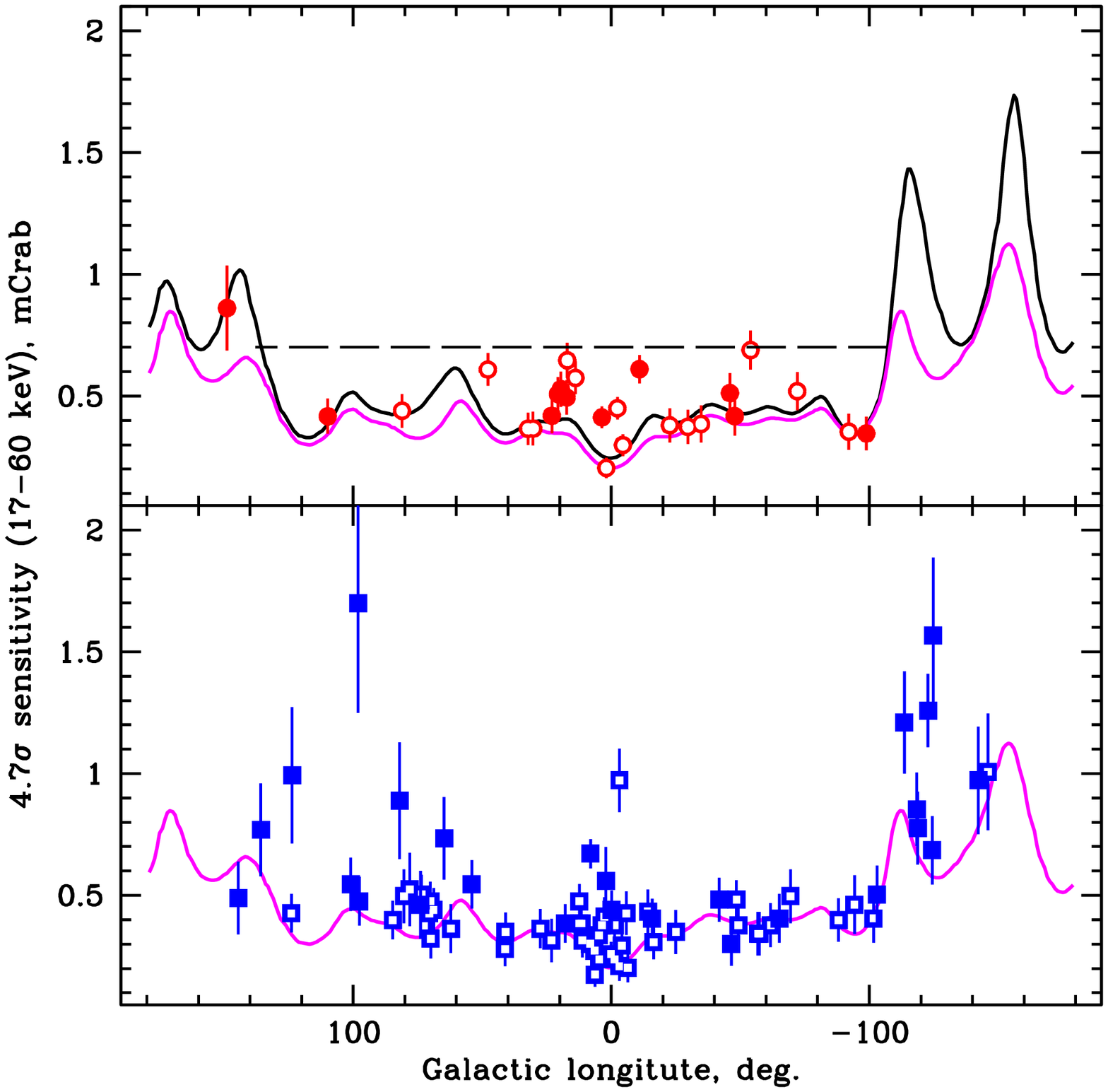}
\caption{Sensitivity of the Galactic plane surveys over the Galactic
  longitude averaged within $|b|<5$\degr in the 17$-$60~keV energy
  band ($4.7\sigma$, 1~mCrab $=1.43\times10^{-11}$\ergscm).  Black
  solid line corresponds to the 9-year survey
  \citet{2012A&A...545A..27K}, magenta line shows the sensitivity
  curve from the current work.  Upper panel: open and filled red
  circles show the positions and fluxes of 26 non-identified
  persistent sources \citep{2012A&A...545A..27K,
    2013MNRAS.431..327L}. Filled circles denote objects identified up
  to now. Long dashed line represents the flux limit used by
  \citet{2013MNRAS.431..327L} to achieve full completeness of the
  survey in the inner part of the Galaxy (there are no unidentified
  sources above the line). Bottom panel: blue open squares denote {46} non-identified sources detected in the current survey. Filled
  squares show {26} sources with tentatively identified
  nature.}\label{fig:aal}
\end{figure}

The \int\ observatory has been successfully operating in orbit since
its launch in October 2002. Over the past years, \int\ acquired a huge
data set, which allowed us to construct high quality X-ray catalogs
in the Galactic Plane (GP), starting from our early papers by
\cite{2004AstL...30..382R,2006AstL...32..145R,2004AstL...30..534M},
to more recent surveys \citep[see][and references therein]{2012A&A...545A..27K,2015MNRAS.448.3766K,2016ApJS..223...15B}.
  These works were subsequently used for many relevant studies,
including systematic discoveries of strongly absorbed high-mass X-ray
binaries (HMXBs) and the study of their luminosity function and
distribution in the Galaxy
\citep{2005A&A...444..821L,2013MNRAS.431..327L,2007A&A...467..585B,2012ApJ...744..108B,2008A&A...484..783C,2013A&A...560A.108C},
the statistics of low mass X-ray binaries \citep{2008A&A...491..209R}
and cataclysmic variables (CVs)
\citep{2008A&A...489.1121R,2010MNRAS.401.2207S}.

In the previous paper \citep{2012A&A...545A..27K}, we presented a
GP survey ($|b| < 17.5$\arcdeg) based on nine years of
\int\ operations. The survey catalogue lists 402 sources detected in
the $17-60$~keV energy band and time-average sky maps at more than
$4.7\sigma$ significance, including 253 Galactic sources of known or
tentatively identified nature, and 34 unidentified sources. The upper
panel of Fig.~\ref{fig:aal} illustrates the limiting flux of the \int\
9-year survey over the Galactic longitude, demonstrating that the survey's
completeness in the inner part of the Galaxy raises to 100\% above the flux limit of 0.7~mCrab (shown
in the figure with the dashed line). The sample of 26 persistent
unidentified sources is shown by red circles \cite[see][for
details]{2013MNRAS.431..327L}.  A number of multi-wavelength follow-up
observations were initiated to unveil the nature of these unclassified objects
\citep{2012AstL...38..629K,2013A&A...556A.120M,2013AstL...39..523R,2015MNRAS.446.2418Z,2013AstL...39..513L,2015AstL...41..179L,2015MNRAS.449..597T,2016MNRAS.460..513T,2016ApJ...816...38T,2016MNRAS.461..304C,2016AstL...42..295B,2017MNRAS.465.1563R}
which led to the classification of 11 sources (shown by solid red circles
in Fig.~\ref{fig:aal}), rising the total survey identification
completeness from $\mysim92\%$ to $\mysim94\%$.

The 9-year \int\ Galactic survey by \cite{2012A&A...545A..27K} and the
all-sky survey by \cite{2016ApJS..223...15B} were based on similar data sets,
available by January 2011 and by the end of 2010, respectively. Over about 6
years that have passed since then, \int\ accumulated an additional
$\mysim80$~Ms and $\mysim50$~Ms of exposure (dead-time corrected) over the
whole sky and in the GP ($|b|<17.5$\degr), respectively. The
increased sensitivity of the currently available \int\ data set allows us to
make a next iteration in the process of finding previously unknown hard X-ray
sources. \cite{2016MNRAS.459..140M} recently released a $17-60$~keV deep
survey of three extragalactic fields (M81, Large Magellanic Cloud and 3C
273/Coma), based on 12 years of observations (2003-2015) with the detection
of 147 sources at $S/N > 4\sigma$, including 37 sources observed in hard
X-rays for the first time.

In this short report we present a catalogue of newly discovered hard
X-ray sources detected in the latest maps of the GP comprising 14
years of data acquired with \int/IBIS.

\begin{figure}
\centerline{
\includegraphics[angle=-90,clip,width=1.4\columnwidth]{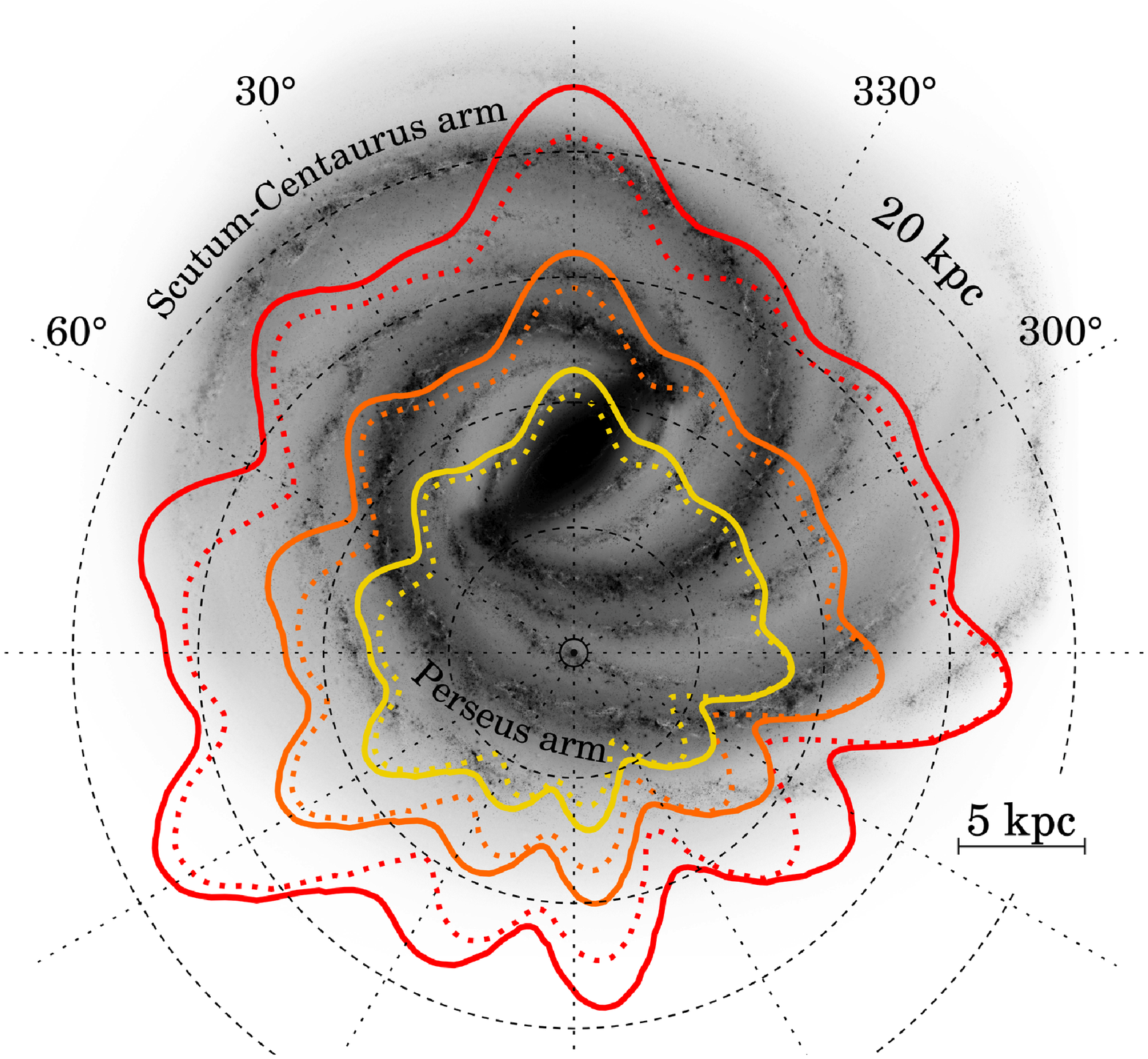}}
\caption{Face-on view of the Galaxy shown along with the distance
  range at which an X-ray source of a given luminosity $L_{\rm HX}$
  (or more) can be detected according to the $17-60$~keV sensitivity
  of the current 14-year \int\ survey (solid lines), compared to the
  9-year GP survey (\citealt{2012A&A...545A..27K}; dotted lines). Red,
  orange and yellow contours correspond to $L_{\rm
    HX}=2\times10^{35}$, $10^{35}$ and $5\times10^{34}$~\lum,
  respectively.  The background image is a sketch of the Galaxy
  adopted from \citet{2009PASP..121..213C}.\label{fig:galaxy}}
\end{figure}

\section{Data analysis}
\label{sec:data}

For this work, we selected all publicly available \int\ data from
December 2002 to March 2017 (spacecraft revolutions 26-1790). Prior to
actual data analysis we applied the latest energy calibration
\citep{2013arXiv1304.1349C} for the registered IBIS/ISGRI detector
events with the \int\ Offline Scientific Analysis version 10.2
provided by ISDC Data Centre for Astrophysics up to the {\sc COR}
level. Then events were processed with a proprietary analysis package
developed at IKI\footnote{Space Research Institute of the Russian
  Academy of Sciences, Moscow, Russia} \citep[details availiable
in][]{2010A&A...519A.107K,2012A&A...545A..27K,2014Natur.512..406C} to
produce a $17-60$~keV sky image of every individual \int\ observation
with a typical exposure time of 2 ks (usually referred as
\textit{Science Window}, or \textit{ScW}). The flux scale in each
\textit{ScW} sky image was renormalized using the flux of the Crab
nebula measured in the nearest observation. This procedure was used to
account for the loss of sensitivity at low energies caused by ongoing
detector degradation.

In total, we obtained 124727 \textit{ScW} images covering the whole
sky, comprising $\mysim220$~Ms of the effective (dead time-corrected)
exposure. For the purposes of this work we selected 79234
\textit{ScW}s ($\mysim130$~Ms) within the GP
($|b|<17.5$\degr). Following \cite{2012A&A...545A..27K} we constructed
six overlapping 70\arcdeg$\times$35\arcdeg cartesian projections
centered at the GP ($|b|=0$\arcdeg) and Galactic
longitudes $l=0^{\circ},\pm50^{\circ},\pm115^{\circ}$, and
$l=180^{\circ}$.

The peak sensitivity of the survey is $2.2\times10^{-12}$ \ergscm\
($\mysim0.15$~mCrab\footnote{~1\,mCrab corresponds to
$1.43\times10^{-11}$~erg~s$^{-1}$~cm$^{-2}$ assuming a spectral shape
$10(E/{\rm 1~keV})^{-2.1}$ photons~cm$^{-2}$~s$^{-1}$~keV$^{-1}$.} in the
17-60~keV energy band) at a $4.7\sigma$ detection level. The survey covers
$90\%$ of the geometrical area ($12680$ degrees) down to the flux limit of
$1.3\times10^{-11}$ \ergscm\ ($\mysim0.93$~mCrab) and $10\%$ of the total
area down to the flux limit of $3.8\times10^{-12}$ \ergscm\
($\mysim0.26$~mCrab). Given the added exposure in the GP, the
achieved improvement in sensitivity with respect to the 9-year survey is in
the range of $10-30\%$. The updated sensitivity of the current survey over
the Galactic longitude is shown in the bottom panel of Fig.~\ref{fig:aal}.
Note that the overall improvement in sensitivity makes it possible to probe
deeper into the Galaxy. Fig.~\ref{fig:galaxy} shows a face-on schematic view
of the Galaxy and the distances at which we can detect a hard X-ray source of
a given luminosity $L_{\rm HX}$ in the $17-60$~keV band. One can see that (i)
we can now detect all sources with the luminosity $L_{\rm
HX}>2\times10^{35}$~\lum at the far end of the Galaxy in the direction
towards the Galactic Centre (GC), (ii) the distance range for the luminosity
$L_{\rm HX}>2\times10^{35}$~\lum covers most of the Galactic stellar mass,
and (iii) the Galactic central bar is fully reachable at luminosities $L_{\rm
HX}>5\times10^{34}$~\lum.

Following \cite{2012A&A...545A..27K}, we adopted a conservative
detection threshold of $(S/N)_{\rm lim}>4.7\sigma$ to ensure that the
final catalogue contains no more than one spurious source assuming
Poisson statistics. The regions around bright sources, such as Crab,
Sco\,X-1, Cyg\,X-1, Cyg\,X-3, Vela\,X-1, GX\,301-2, and GRS\,1915+105 were
excluded from the automated excess selection to prevent false
detections triggered by high systematic noise. However, manual
inspection of these regions was performed to select possible source
candidates (properly marked as being detected in noisy environment).

A special care was taken for the source detection in the region of
$\mysim17$ degrees around the GC due to enchanced systematics
\citep[see e.g.][]{2010A&A...519A.107K}. False detections were
revealed by a distorted excess shape that differs significantly from
the instrumental point-spread function, which is a symmetric
two-dimensional Gaussian ($\sigma=5$\arcmin). One can reduce
IBIS/ISGRI false detections by using additional information from the BAT
coded-mask telescope \citep{2005SSRv..120..143B} onboard
the \textit{Swift} observatory \citep{2004ApJ...611.1005G} working at hard
X-ray energies. Since BAT has a different coded-mask design
compared to IBIS, it suffers different systematics, which allows one
to supress the non-statistical uncertainties known to IBIS \citep[the
  idea on which the combined \textit{Swift}-\int\ survey
by][is based]{2012ApJS..201...34B}. We assume that finding a hard X-ray
counterpart in the ongoing \textit{Swift}/BAT surveys
\citep{2010A&A...524A..64C,2013ApJS..207...19B} of a suspected
IBIS/ISGRI systematic excess adds more evidence that the excess is a
real source.

\section{Results}

\begin{table*}
\noindent
\centering
\caption{The list of known X-ray transients detected in 14-year
time averaged map at $S/N>4.7\sigma$ mainly due to outburst event(s)
occured between 2010 and 2016.}\label{tab:trans} 
\centering
\vspace{1mm}
  \begin{tabular}{|c|l|c|c|c|r|c|c|l|l|}
\hline\hline
No. & Name & \multicolumn{1}{c}{RA (J2000)} & \multicolumn{1}{c}{Dec (J2000)} & Flux$_{\rm 17-60~keV}$ & $S/N$ & Type$^{a}$ & Outburst \\ 
      &  & \multicolumn{1}{c}{(deg)} & \multicolumn{1}{c}{(deg)} & $10^{-11}$\ergscm & & &  year\\
\hline

1  & GS 0834-430 & 128.979 & -43.185 & $0.99 \pm 0.08$ & 12.6  & HMXB& 2012\\
2  & GS 1354$-$64 & 209.562 & -64.733 & $3.94 \pm 0.09$ & 45.2&LMXB& 2015\\
3  & IGR J17177$-$3656 & 259.424 & -36.880 & $0.32 \pm 0.06$ & 4.9  &LMXB&  2011  \\
4  & GRS 1716$-$249 & 259.903 & -25.020 & $4.03 \pm 0.06$ & 65.2  &LMXB& 2016\\
5   &Swift J1734.5$-$3027 & 263.652 & -30.399 & $0.87 \pm 0.05$ & 17.2& LMXB& 2013\\
6  & GRS 1739$-$278 & 265.661 & -27.748 & $2.11 \pm 0.05$ & 42.5&LMXB& 2014, 2016\\
7  & GRO J1744$-$28 & 266.138 & -28.741 & $4.51 \pm 0.05$ & 91.8&LMXB& 2014, 2017\\
8  & Swift J174510.8$-$262411 & 266.297 & -26.401 & $30.74 \pm 0.05$ &607.7  &LMXB&  2012\\
9  & IGR J17498$-$2921 & 267.482 & -29.323 & $0.72 \pm 0.05$ & 14.6  &  LMXB & 2011\\
10  & 1RXS J180408.9$-$342058 & 271.036 & -34.356 & $5.80 \pm 0.06$ &99.5  &LMXB&  2012\\
11  & SAX J1806.5$-$2215 & 271.634 & -22.233 & $3.31 \pm 0.06$ & 54.7&LMXB& 2011\\
12  & IGR J18179$-$1621 & 274.477 & -16.481 & $0.38 \pm 0.08$ & 4.7&HMXB&2012\\
13  & IGR J18245$-$2452 & 276.106 & -24.879 & $1.29 \pm 0.07$ & 19.3&LMXB& 2013\\
14  & MAXI J1828$-$249 & 277.238 & -25.041 & $1.23 \pm 0.07$ & 17.3&BHC& 2013\\
15  & MAXI J1836$-$194 & 278.937 & -19.314 & $2.17 \pm 0.09$ & 24.5&XRB& 2011 \\
16  & XTE J1859+083  &284.753&8.239 &$0.60\pm 0.07$ & 8.3&HMXB&2015\\
17  & V404 Cyg & 306.019 & 33.867 & $9.70 \pm 0.08$ & 119.8  &LMXB&2015 \\

\hline

\end{tabular}\\
\begin{flushleft}
$^{a}$ General astrophysical type of the object: LMXB (HMXB) -- low-
(high-) mass X-ray binary; BHC -- black hole candidate;  XRB -- X-ray binary.
\end{flushleft}
\vspace{3mm}
\end{table*}


Our analysis of 14-year averaged sky images of the GP
($|b|<17.5$\degr) led to the detection of {522} hard X-ray sources at
significance $S/N>4.7\sigma$, which is $\mysim$30\% more compared to 402
sources detected in the 9-year survey \citep{2012A&A...545A..27K} with the
same detection threshold. Note that 14 weak sources\footnote{IGR
J17315$-$3221, IGR J17331$-$2406, Swift J2113.5+5422, IGR J18175$-$1530, IGR
J17448$-$3231, XTE J1543$-$568, IGR J16293$-$4603, IGR J17197$-$3010, IGR
J16358$-$4726, IGR J18497$-$0248, XTE J1751$-$305, AX J1753.5$-$2745, IGR
J09189$-$4418 and IGR J20107+4534.} listed in the 9-year survey with fluxes
of $0.2-0.5$~mCrab are not detected in the current study, probably due to an
intrinsic variability. Among {134} newly added sources {(522-402+14)}, we identified {62} previously known X-ray emitters,
including 17 known sources that experienced transient events after 2010
(Table~\ref{tab:trans}).  A detailed analysis of the survey's catalogue will
be presented elsewhere. For the current report we selected those {72}
(out of {134}) newly detected hard X-ray sources that have not been
listed in the \int\ surveys based on the data acquired before 2010
\citep{2007A&A...475..775K,2010A&A...523A..61K,2012A&A...545A..27K,2004ApJ...607L..33B,2006ApJ...636..765B,2007ApJS..170..175B,2010ApJS..186....1B,2016ApJS..223...15B},
i.e. those sources whose detection is mainly determined by the $\mysim$6-year
increased \int\ survey sensitivity.

Table~2\footnote{Table~2 is only available in the online version of
  the paper.} lists new \int\ sources detected in the current work
with significances between $4.7\sigma$ and $15\sigma$ and fluxes
between 0.17 and 1.7 mCrab
($2.5\times10^{-12}-2.4\times10^{-11}$\ergscm). We searched for source
counterparts within a 3.6$'$ error circle (90\% confidence), as
typical for the \int\ sources detected at $S/N=5-6\sigma$
\citep{2007A&A...475..775K}. As seen from Table~2, {31} source
candidates are also detected in the ongoing \textit{Swift}/BAT all-sky
hard X-ray survey \citep{2010A&A...524A..64C,2013ApJS..207...19B}. No
any hard X-ray counterpart were found for {41} sources, thus they have
been detected in hard X-rays for the first time.

We utilized also the
SIMBAD\footnote{\url{http://simbad.u-strasbg.fr/simbad}} and
NED\footnote{\url{http://ned.ipac.caltech.edu}} data bases to perform
a preliminary identification of the detected source candidates within
3.6$'$ of the \int\ position. However, usually unique optical/IR
counterparts and hence firm astronomical classification can only be
obtained based on arcsecond positions provided by soft X-ray focusing
telescopes. Therefore, we paid a special attention for finding soft
X-ray counterparts in the
HEASARC\footnote{\url{https://heasarc.gsfc.nasa.gov}} data base,
\textit{Swift}/XRT point source catalogue
\citep[1SXPS;][]{2014ApJS..210....8E} and the third \text{XMM-Newton}
serendipitous source catalogue
\citep[3XMM-DR5;][]{2016A&A...590A...1R}. As a result, we suggest
classification for {26} sources from the list, with two
(IGR~J00555+4610 and IGR~J18184$-$2352) being most likely CVs and {21}
probably being active galactic nuclei (Table~2 and filled squares in
the bottom panel of Fig.\,\ref{fig:aal}). The remaining {46}
unclassified sources are shown by open squares in the bottom panel of
Fig.~\ref{fig:aal}.  Note, that most of them are detected close to the
limiting flux of the survey below $\mysim0.5$~mCrab (except for
IGR~J16459$-$2325 with a measured flux of $1.0\pm0.1$~mCrab). {Twenty}
out of the {46} nonidentified sources are located in the Galactic
bulge at $|l|<15$\degr.

\section{Concluding remark}

Regular observations of the GP with \int\ are consistently
improving the sensitivity of the hard X-ray survey and allowing us to
extend our knowledge of the Galactic X-ray source population, both
for weak and nearby sources \citep[mostly CVs, see
e.g.][]{2010AstL...36..904L,2016MNRAS.461..304C,2016MNRAS.460..513T},
and more distant objects located at far end of the Galaxy
\citep{2016MNRAS.462.3823L,2017MNRAS.465.1563R}. The presented
catalogue opens the path to a large program of follow-up observations,
dedicated both to unveil new classes of objects and to increase the
overall completeness of the source sample, needed for many Galactic
population studies.

\section*{Acknowledgments}

This work is based on observations with \int, an ESA project with
instruments and the science data centre funded by ESA member states
(especially the PI countries: Denmark, France, Germany, Italy,
Switzerland, Spain), and Poland, and with the participation of Russia
and the USA. This research has made use of: data obtained from the
High Energy Astrophysics Science Archive Research Center (HEASARC)
provided by NASA's Goddard Space Flight Center; the SIMBAD database
operated at CDS, Strasbourg, France; the NASA/IPAC Extragalactic
Database (NED) operated by the Jet Propulsion Laboratory, California
Institute of Technology, under contract with the National Aeronautics
and Space Administration; the Palermo BAT Catalogue and database
operated at INAF -- IASF Palermo. The data were obtained from the
European\footnote{\url{http://isdc.unige.ch}} and
Russian\footnote{\url{http://hea.iki.rssi.ru/rsdc}} \int\ Science Data
Centers. The authors are grateful to E.M. Churazov, who developed the
INTEGRAL/IBIS data analysis methods and provided the software, and
thank the Max Planck Institute for Astrophysics for computational
support. This work was financially supported by Russian Science
Foundation grant 14-22-00271.

\bibliographystyle{mnras}
\bibliography{references}

{\small

\onecolumn
\begin{landscape}
\begin{minipage}{0.8\textwidth}
{\normalsize {\bf Table 2.} The list of newly detected \int\ hard X-ray sources based on 14 years of observations. This catalogue is only available in the online version of the paper.}
\end{minipage}
\begin{longtable}{clrrrrcclll}

\hline 
\multicolumn{1}{|c|}{No.} &
\multicolumn{1}{c|}{Name\footnotemark[1]} & 
\multicolumn{1}{c|}{RA\footnotemark[2]} & 
\multicolumn{1}{c|}{Dec\footnotemark[2]} &
\multicolumn{1}{c|}{Flux\footnotemark[3]} &
\multicolumn{1}{c|}{S/N} &
\multicolumn{1}{c|}{Type\footnotemark[4]} &
\multicolumn{1}{c|}{Ref\footnotemark[5]} &
\multicolumn{1}{c|}{Hard X-ray/$\gamma$} &
\multicolumn{1}{c|}{Soft X-ray} &
\multicolumn{1}{c|}{Optical/IR} \\

\multicolumn{1}{|c|}{} &
\multicolumn{1}{c|}{} &
\multicolumn{1}{c|}{(deg)} &
\multicolumn{1}{c|}{(deg)} &
\multicolumn{1}{c|}{(17--60~keV)} &
\multicolumn{1}{c|}{} &
\multicolumn{1}{c|}{} &
\multicolumn{1}{c|}{} &
\multicolumn{1}{c|}{counterpart} &
\multicolumn{1}{c|}{counterpart} &
\multicolumn{1}{c|}{counterpart} \\ \hline 
\endfirsthead

\multicolumn{11}{c}%
{{\bfseries Table 2  -- continued from previous page}} \\
\hline 
\multicolumn{1}{|c|}{No.} &
\multicolumn{1}{c|}{Name\footnotemark[1]} &          
\multicolumn{1}{c|}{RA\footnotemark[2]} &                     
\multicolumn{1}{c|}{Dec\footnotemark[2]} &
\multicolumn{1}{c|}{Flux\footnotemark[3]} &
\multicolumn{1}{c|}{S/N} &
\multicolumn{1}{c|}{Type\footnotemark[4]} &
\multicolumn{1}{c|}{Ref.\footnotemark[5]} &
\multicolumn{1}{c|}{Hard X-ray/$\gamma$} &
\multicolumn{1}{c|}{Soft X-ray} &
\multicolumn{1}{c|}{Optical/IR} \\

\multicolumn{1}{|c|}{} &
\multicolumn{1}{c|}{} &
\multicolumn{1}{c|}{(deg)} &
\multicolumn{1}{c|}{(deg)} &
\multicolumn{1}{c|}{(17--60~keV)} &
\multicolumn{1}{c|}{} &
\multicolumn{1}{c|}{} &
\multicolumn{1}{c|}{} &
\multicolumn{1}{c|}{counterpart} &
\multicolumn{1}{c|}{counterpart} &
\multicolumn{1}{c|}{counterpart} \\ \hline 
\endhead

\hline \multicolumn{11}{|r|}{{Continued on next page}} \\ \hline
\endfoot

\hline \hline
\endlastfoot

1  & J00555+4610  & 13.876 & 46.173 & $1.42\pm 0.28$&5.1& CV &1,2 &Swift J0055.4+4612 & XSS J00564+4548& \\
2 & J01017+6519    & 15.440   & 65.330  & $0.61 \pm 0.08$ & 8.0 & & & & \\
3  & J02145+5142  & 33.625 & 51.710 &$1.10 \pm 0.19$ &5.7& AGN &1 &  Swift J0213.7+5147& &\\
4  & J03117+5028 & 47.934 & 50.472 & $0.70\pm 0.15$ & 4.7  & AGN&1 &Swift J0311.9+5032 & 1RXS J031202.7+502922 & 2MASX J03120291+5029147 \\
5  & J06402$-$2552  & 100.050 & -25.868 & $2.24 \pm 0.32$ & 7.0  &AGN& 1&Swift J0640.4$-$2554&&ESO 490$-$26\\
6 & J07141+0146    & 108.547 & 1.744    &  $1.44 \pm 0.24$ & 6.0  & & & &\\
7  & J07258+0054 & 111.471 & -0.907 & $1.39 \pm 0.22$ &6.3&AGN&1&Swift J0725.7$-$0055&&QSO B0723$-$007\\
8  & J07396$-$3143& 114.908 & -31.732 & $1.73 \pm 0.21$ & 8.1&AGN?&1&Swift J0739.7$-$3142&&2MASX J07394469$-$3143024\\
9  & J07433$-$2544& 115.844 & -25.736 & $1.22 \pm 0.15$ & 7.9&AGN?&1&Swift J0743.3$-$2546& 1RXS J074315.6$-$254545& 2MASX J07431472$-$2545501 \\
10  & J08004$-$4309 & 120.121 & -43.160 & $0.58 \pm 0.10$ & 5.7 & &1&Swift J0800.7$-$4309&&\\
11 & J08215$-$1320 & 125.387 & -13.339 & $0.98 \pm 0.14$ & 6.9  &AGN?&3 & && NGC 2578, PGC023449\\

12 & J08321$-$1808 & 128.022 & -18.141 & $1.11 \pm 0.15$ & 7.3  &
AGN?& 4,5,6 & & $\bfrac{\rm 1RXS J083158.1-180828}{\rm 1SXPS~J083158.6-180840}$ & \\
13 & J08398$-$1214 & 129.961 & -12.243 & $1.80 \pm 0.15$ & 11.7 &AGN? & 5,1& Swift J0839.6$-$1213&1RXS J083950.7$-$121424 & 2E 2028 \\
14  &  J08453$-$3529 & 131.329 & -35.493 & $0.72 \pm 0.12$ & 6.0  & AGN&1,7& Swift J0845.0$-$3531&	1RXS J084521.7$-$353048 & WISE J084521.37$-$353024.2 \\
15 & J09278$-$3935 & 141.965 & -39.590 & $0.66 \pm 0.12$ & 5.7  & & & & \\
16 & J09331$-$4725 & 143.302 & -47.441 & $0.57 \pm 0.09$ & 6.2  & & && \\
17    & J11275$-$5319 & 171.898 & -53.369 & $0.71 \pm 0.11$ & 6.5  &&8 & PBC J1127.7$-$5320 &&  \\
18 & J11299$-$6557 & 172.490  & -65.960  &  $0.58 \pm 0.10$ & 5.8 &AGN?&5  & & 1RXS J112955.1$-$655542& 2MASS J11295643$-$6555218 \\
19 & J12086$-$6327 & 182.157 & -63.452 &  $0.54 \pm 0.09$ & 6.0  & & & &\\
20  & J12489-5930 & 192.161 & -59.507 & $0.49 \pm 0.09$ & 5.5  & & & & &\\
21 & J12529$-$6351 & 193.241 & -63.868 & $0.49 \pm 0.09$ &5.5  & &  & &\\
22  & J14044$-$6146 & 211.029 & -61.700 & $0.69 \pm 0.08$ & 8.1  & & 1&  Swift J1403.6$-$6146 && \\
23 & J14417$-$5533 & 220.427 & -55.550 & $0.69 \pm 0.09$ & 7.4   &AGN?&5 & & 1RXS J144116.4$-$553329 & 2MASS J14411645$-$5533306 \\
24  & J14192$-$6048  & 214.819 & -60.810 & $0.43 \pm 0.09$ & 5.1  &PSR/PWN&1&Swift J1418.8$-$6055&AX J1418.7$-$6058& \\
25  & J15550$-$4306 & 238.769 & -43.090 & $0.50 \pm 0.09$ & 5.7  & & & && \\
26 & J16459$-$2325 & 251.477 & -23.428 & $1.39 \pm 0.13$ & 11.0  & & && \\
27 & J16494$-$1740 & 252.355 & -17.676 & $0.80 \pm 0.14$ & 5.6  &AGN? &4,9 & 4PBC J1649.3$-$1738 & 1SXPS J164920.9$-$173840 & ESO 586$-$4 \\
28  & J17040$-$4305 & 256.010 & -43.080 & $0.44 \pm 0.07$ & 6.2  & & & &  1RXS J170406.3$-$430637&\\
29 & J17098$-$2344 & 257.455 & -23.747 & $0.63 \pm 0.08$ & 8.4  &AGN?&5,6 & 4PBC J1709.7$-$2348 & 1RXS J170944.9$-$234658  & 2MASS J17094469$-$2346531\\
30 & J17158$-$2124 & 258.959 & -21.411 & $0.59 \pm 0.08$ & 7.5  & & && \\
31  & J17255$-$4509 & 261.380 & -45.170 & $0.58 \pm 0.08$ &7.1  & AGN? & 8,9& 4PBC J1725.8$-$4510 & &2MASX J17253053$-$4510279\\
32  & J17326$-$3445 & 263.169 & -34.754 & $0.29 \pm 0.06$ & 5.2  & & & &1RXS J173251.1$-$344728&\\
33  & J17327$-$4405 & 263.183 & -44.103 & $0.62 \pm 0.09$ & 6.9  & & & &\\
34  & J17422$-$2108     & 265.560 & -21.106 &$0.25 \pm 0.05$ &  5.0 & &6 & & 1SXPS J174211.7$-$210354& \\ 
35  & J17528$-$2022 & 268.217 & -20.367 & $0.43 \pm 0.06$ &6.7  & & 8 & 4PBC J1752.6$-$2020&&\\
36  & J17538$-$2544 & 268.463 & -25.749 & $0.54 \pm 0.05$ & 10.5  & & 10 & Swift J1753.7$-$2544 &&  \\
37  & J17570$-$2500 & 269.266 & -25.015 & $0.34 \pm 0.05$ & 6.4  & & && \\
38  & J17596$-$2315 & 269.907 & -23.266 &  $0.39 \pm 0.06$ & 6.9  & & & &\\
39  & J18010$-$3045 & 270.271 & -30.764 & $0.37 \pm 0.05$ & 7.2  & & && \\
40  & J18013$-$3222 & 270.326 & -32.371 & $0.34 \pm 0.05$ & 6.4  & & && \\
41  & J18017$-$3542 & 270.371 & -35.638 & $0.42 \pm 0.06$ &7.0  & &8 & 4PBC J1801.7$-$3540 &\\
42  & J18044$-$1829 & 271.107 & -18.487 & $0.45 \pm 0.07$ & 6.3  & & & &\\
43  & J18070$-$3507 & 271.750 & -35.132 & $0.30 \pm 0.06$ & 4.9  & & & && \\
44  & J18102$-$1751 & 272.555 & -17.853 & $0.55 \pm 0.07$ & 7.4  & & && \\
45  & J18112$-$2641 & 272.854 & -26.707 & $0.48 \pm 0.06$ & 8.7  & & & && \\
46  & J18141$-$0606 & 273.525 & -6.114   & $0.45 \pm 0.09$ & 5.1  & & && \\
47  & J18141$-$1823 & 273.538 & -18.395 & $0.68 \pm 0.07$ &9.0  & &8 & 4PBC J1814.1$-$1822 &\\
48  & J18147$-$3400 & 273.690  & -34.010   &   $0.54 \pm 0.06$ & 8.6  && &&\\
49  & J18165$-$3912 & 274.136 & -39.202 &  $0.61 \pm 0.09$ & 7.0   & & && \\
50  & J18172$-$1944 & 274.307 & -19.740 & $0.55 \pm 0.07$ & 7.6  & &  && \\
51  &  J18184$-$2352& 274.610  &  -23.880  & $0.96 \pm 0.06$ & 15.2  & CV? & 11 & &MACHO 311.37389.3983& \\
52  & J18263$-$1345 & 276.576 & -13.753 & $0.55 \pm 0.08$ & 6.6  & PWN? && HESS J1825$-$137? && \\
53  & J13545$-$5958 & 208.621 & -59.982 & $0.54 \pm 0.09$ & 6.4  & & & && \\
54  & J18434$-$0508 & 280.855 & -5.138   & $0.52 \pm 0.08$ & 6.2 & & 8,12 & 4PBC J1842.8$-$0506 & Swift J184311.0$-$050539& \\
55  & J18544+0839$\dagger$     & 283.605 & 8.661 &  $0.50 \pm 0.08$ & 6.7  & & &&\\
56  & J19039+3348    &  285.948& 33.825   & $1.05 \pm 0.17$ &6.2 & AGN &1 & Swift J1903.9+3349& \\
57  & J19071+0716    & 286.783 & 7.274    & $0.40 \pm 0.07$ & 5.7 &&5,13& & $\bfrac{\rm 1SXPS~J190706.3+072004}{\rm 3XMM~J190706.3+072003}$ &  \\
58  & J19260+4136 & 291.517 & 41.608 & $0.66 \pm 0.14$ & 4.7  & AGN&1&  Swift J1926.9+4140& 1RXS J192630.6+413314 & 2MASX J19263018+4133053\\
59 & J19305+1851 & 292.632 & 18.857 & $0.78\pm0.10$ &7.7&SNR?&&& 1RXS J193029.9+185205& \\
60  & J19421+3613    & 295.530 & 36.219  & $0.61 \pm 0.09$ & 6.6  & & & & \\
61  & J19498+2534    & 297.470 & 25.557  & $0.52 \pm 0.10$ &5.0  & &8,14 & 4PBC J1950.0+2532&AX J1949.8+2534 &\\
62  & J19577+3339 & 299.429 & 33.658 & $0.46 \pm 0.08$ & 5.6  & & & && \\
63  & J19504+3318$\dagger$     & 297.615& 33.311 & $0.63 \pm 0.09$ & 7.4  & & &  & 1RXS J195020.5+331419&\\
64  & J20063+3641    & 301.601 & 36.684  & $0.72 \pm 0.08$ &9.1  & &1 & Swift J2006.4+3645&& \\
65  & J20084+3221    & 302.124 & 32.350  & $0.68 \pm 0.08$ & 8.4 & &5,8 &4PBC J2008.7+3221 &1SXPS J200843.8+321824 & \\
66  & J20596+4303  & 314.914 & 43.054 & $0.57 \pm 0.08$ & 6.8  & &1&  Swift J2059.6+4301A/B& &\\
67  & J21099+3533 & 317.490 & 35.560 & $0.71 \pm 0.11$ & 6.6  & & & &&\\
68  & J21133+3154 & 318.328 & 31.923 & $0.75 \pm 0.15$ & 4.8  & & & & 1RXS J211319.3+315211&\\
69  & J21382+3204 & 324.563 & 32.072 & $1.27 \pm 0.24$ & 5.4  &AGN&1&Swift J2138.8-3207&	1RXS J213833.0+320507 & WISE J213833.43+320505.8\\
70 & J21397+5949 & 324.945 & 59.832 & $0.78 \pm 0.11$ & 7.1  &AGN&1& Swift J2139.7+5951& 1RXS J213944.3+595016 & WISE J213944.96+595015.1\\
71  & J22018+5049  & 330.474 & 50.830 & $0.68 \pm 0.10$ & 6.7  &AGN&1&  Swift J2201.9+5057 &&87GB 215950.2+503417\\
72  & J22455+3940  & 341.383 & 39.683 & $2.43 \pm 0.45$ & 5.4  & AGN&1&  Swift J2246.0+3941 &&3C452\\

\footnotetext[1]{{\it INTEGRAL} (IGR) name of the source. A dagger
  symbol $\dagger$ marks that the source is located in the region of
  high systematic noise, and that its measured flux should be taken
  with the caution.}  \footnotetext[2]{Equatorial coordinates (right
  ascension and declination) are in standard J2000.0 epoch.}
\footnotetext[3]{The measured 17--60~keV flux of the source
  $\times10^{-11} \textrm{erg~} \textrm{cm}^{-2}\textrm{s}^{-1}$.}
\footnotetext[4]{General astrophysical type of the object: AGN --
  active galactic nucleus, SNR -- supernova remnant; CV -- cataclysmic
  variable; PSR -- isolated pulsar or pulsar wind nebula (PWN). A
  question mark indicates that the specified type should be
  confirmed.}  \footnotetext[5]{References. -- (1)
  \cite{2013ApJS..207...19B}; (2) \cite{2006AstL...32..588B}; (3)
  \cite{2003A&A...412...45P}; (4) \cite{2013ApJS..207...16M}; (5)
  \cite{2012ApJ...751...52E}; (6) \cite{2014ApJS..210....8E}; (7)
  \cite{2010A&A...519A..96M}; (8) \cite{2010A&A...524A..64C}; (9)
  \citet{2006AJ....131.1163S,2012ApJS..199...26H}; (10)
  \cite{2013ApJS..209...14K}; (11) \cite{2004PASP..116..610C}; (12)
  \cite{2013ATel.5200....1R}; (13) \cite{2016A&A...590A...1R}; (14)
  \cite{2015ATel.8250....1S,2001ApJS..134...77S}; }
\end{longtable}
\end{landscape}

}
\label{lastpage}

\end{document}